\documentclass[prd,amssymb,amsmath,twocolumn,showpacs]{revtex4}
\usepackage{epsfig}
\newcommand{\beq}{\begin{equation}}
\newcommand{\eeq}{\end{equation}}
\newcommand{\bqa}{\begin{eqnarray}}
\newcommand{\eqa}{\end{eqnarray}}
\newcommand{\fr}{\frac}

\begin{document}
\title{No-horizon theorem for vacuum gravity with spacelike $G_{1}$ isometry groups}
\author{S\'{e}rgio M. C. V. Gon\c{c}alves}
\affiliation{Department of Physics, Yale University, New Haven, Connecticut 06511}
\date{\today}
\begin{abstract}
We show that $(3+1)$ vacuum spacetimes admitting a global, spacelike, one-parameter Lie group of isometries of translational type cannot contain apparent horizons.
The only assumption made is that of the existence of a global spacelike Killing vector field with infinite open orbits; the four-dimensional vacuum spacetime metric is otherwise arbitrary. This result may thus be viewed as a hoop conjecture theorem for vacuum gravity with one spacelike translational Killing symmetry.
\end{abstract}
\maketitle

The main purpose of this paper is to probe the relation between geometry and the existence---or lack thereof---of horizons in general relativity. Curiously, this aspect is normally overlooked in favor of the relation between gravitational mass and horizon formation, with the notable exception of the so-called hoop conjecture, which makes a definite (albeit loosely defined) statement about the combined role of geometry and mass in horizon formation: ``Horizons form when and only when a mass $m$ gets compacted into a region whose circumference in {\em every} direction is $C\lesssim4\pi m$''~\cite{thorne72}. Despite inherent ambiguities in the definitions of horizon, mass, and circumference, no known counterexample appears to exist~\cite{nocounterhc}. While many examples corroborating the validity of the conjecture are known, none of these answers the fundamental question of {\em why is it that mass needs to be compacted in all three spatial directions to form a horizon?}

Thus motivated, we investigate here a large class of spacetimes which, by construction, cannot admit spatially bounded distributions of mass, and show that apparent horizons (outer marginally trapped surfaces, which are the outer boundary of a trappped region) cannot develop in such spacetimes. To this end, we consider four-dimensional vacuum spacetimes $(M,{\mathbf g})$, with the minimal assumption that they admit a global spacelike Killing vector field (KVF) of translational type with open orbits, i.e., there is a one-dimensional Lie group of isometries $G_{1}={\mathbb R}$ acting on a three-dimensional submanifold ${\mathcal M}$, such that $M\approx{\mathbb R}\times{\mathcal M}$. This translational symmetry mollifies the restrictions of cylindrical symmetry, which requires an additional KVF with closed orbits (commuting with the $G_{1}$ KVF, such that the orthogonal subspace is integrable), and is given by a two-parameter spacelike Lie group $G_{2}={\mathbb R}\times U(1)$ acting on ${\mathbb R}^{3}$.

It is well known that, in the presence of one global spacelike KVF, Einstein's equations for $(3+1)$ vacuum gravity are equivalent to $(2+1)$ gravity coupled to matter fields related to the norm and twist of the isometry-generating KVF~\cite{geroch71-72,moncrief86}. By studying the dimensionally reduced system induced by the $G_{1}$ group, we show that the three-dimensional matter content obeys the dominant energy condition, which in turn enforces---together with the field equations---the absence of apparent horizons in the $(2+1)$ spacetime. That no apparent horizons form in the full four-dimensional spacetime then follows from its $G_{1}$-induced topological product structure. Natural geometrized units, in which $8\pi G=c=1$, are used throughout.

Proposition 1: {\em Let $(M,\mathbf{g})$ be a $(3+1)$-dimensional vacuum spacetime, admitting a global spacelike one-parameter Lie group of isometries, $G_{1}={\mathbb R}$. Then the Einstein equations for $(M,\mathbf{g})$ are equivalent to those for $(2+1)$ gravity coupled to divergence-free harmonic map fields.}

Proof. In what follows, we adopt the dimensional reduction approach developed by Moncrief~\cite{moncrief86} for spacelike $U(1)$ isometry groups. Let the coordinates in $M$ be $\{x^{3},x^{i}; i=0,1,2\}$, and take the global spacelike KVF to be $\partial_{x^{3}}$, whose space of orbits, under $G_{1}$ actions, induces a three-manifold ${\mathcal M} \approx M/{\mathbb R}$ . The four-metric in $M$ can then be written as
\beq
ds^{2}=e^{-2\phi}\gamma_{ij}dx^{i}dx^{j}+e^{2\phi}(dx^{3}+\beta_{a}dx^{a}+\beta_{0}dt)^{2}, \label{msplit}
\eeq
where $|\partial_{x^{3}}|\equiv e^{\phi}$, and the induced Lorentzian metric on the quotient manifold ${\mathcal M}\approx{\mathbb R}\times\Sigma$ admits the ADM decomposition
\beq
\gamma_{ij}dx^{i}dx^{j}=-\tilde{N}dt^{2}+\tilde{\sigma}_{ab}(dx^{a}+\tilde{N}^{a}dt)(dx^{b}+\tilde{N}^{b}dt), 
\eeq
where the indices $(a,b,c,...)$ refer to two-dimensional quantities, denoted by a tilde, defined on $\Sigma$. Introducing momenta $(\tilde{p}, \tilde{e}^{a}, \tilde{\pi}^{ab})$ conjugate to $(\phi, \tilde{\beta}_{a}, \tilde{\sigma}_{ab})$ in the usual way, the Einstein-Hilbert action is
\bqa
S&=&\int_{\mathcal M} dt d^{2}x (\tilde{\pi}^{ab}\tilde{\sigma}_{ab,t}+\tilde{e}^{a}\tilde{\beta}_{a,t}+\tilde{p}\phi_{,t}-\tilde{N}\tilde{\mathcal H}-\tilde{N}^{a}\tilde{\mathcal H}_{a} \nonumber \\
&&-\beta_{0}\tilde{e}^{a}_{,a}), \label{eha}
\eqa
where the canonical Hamiltonian scalar and momentum vector densities are, respectively,
\bqa
\tilde{\mathcal H}&=&\fr{1}{\sqrt{\tilde{\sigma}}}[\tilde{\pi}^{ab}\tilde{\pi}_{ab}-(\tilde{\pi}^{a}_{a})^{2}+\fr{1}{8}\tilde{p}^{2}+\fr{1}{2}e^{-\phi}\tilde{\sigma}_{ab}\tilde{e}^{a}\tilde{e}^{b}] \nonumber \\
&&+\sqrt{\tilde{\sigma}}\{-^{(2)}\!R+2\tilde{\sigma}^{ab}\phi_{,a}\phi_{,b} \nonumber \\
&&+e^{4\phi}\tilde{\sigma}^{ac}\tilde{\sigma}^{bd}\tilde{\beta}_{[a,b]}\tilde{\beta}_{[c,d]}\}, \\
\tilde{\mathcal H}_{a}&=&-2\tilde{\nabla}_{b}\tilde{\pi}^{b}_{a}+\tilde{p}\phi_{,a}+2\tilde{e}^{b}\tilde{\beta}_{[b,a]}.
\eqa
The constraint equations for the action $S$ are
\beq
\tilde{\mathcal H}=0, \;\;\;\; \tilde{\mathcal H}_{a}=0, \;\;\;\; \tilde{e}^{a}_{,a}=0,
\eeq
and are equivalent to the four-dimensional constraints, restricted to the assumed symmetry class. The ``electromagnetic'' constraint $\tilde{e}^{a}_{,a}=0$ allows for the introduction of a pseudoscalar function $\omega$ (the ``twist potential'') via
\beq
\tilde{e}^{a}:=\epsilon^{ab}\omega_{,b},
\eeq
(we remark that there is in general an additional ``harmonic'' term, which vanishes for ${\mathbb R}^{2}$ topologies) which can be used to replace the explicit dependence of $S$ on $(\tilde{\beta}^{a},\tilde{e}^{a})$ by that on $(\omega,\tilde{r})$, where $\tilde{r}\equiv(\sqrt{\tilde{\sigma}}/\tilde{N})(\omega_{,t}-\tilde{N}^{a}\omega_{,a})$ is the canonical conjugate momentum to $\omega$. In terms of these new variables, Eq. (\ref{eha}) reads
\beq
S=\int_{\mathcal M} dt d^{2}x(\tilde{\pi}^{ab}\tilde{\sigma}_{ab,t}-\tilde{N}\tilde{\mathcal H}_{\rm G}-\tilde{N}^{a}\tilde{\mathcal H}^{\rm G}_{a}) + S_{\rm M}, \label{eha2}
\eeq
where the first integral is just the canonical action for pure $(2+1)$ gravity, and 
\bqa
S_{\rm M}&=&\int_{\mathcal M} dtd^{2}x\left\{\tilde{p}\phi_{,t}+\tilde{r}\omega_{,t}-\tilde{N}\left[\fr{1}{\sqrt{\tilde{\sigma}}}\left(\fr{\tilde{p}^{2}}{8}+\fr{e^{4\phi}}{2}\tilde{r}^{2}\right) \right. \right. \nonumber \\
&&\left.+\sqrt{\tilde{\sigma}}(2\tilde{\sigma}^{ab}\phi_{,a}\phi_{,b}+\fr{1}{2}e^{-4\phi}\tilde{\sigma}^{ab}\omega_{,a}\omega_{,b})\right] \nonumber \\
&&-\tilde{N}^{a}(\tilde{p}\phi_{,a}+\tilde{r}\omega_{,a})\} \nonumber \\
&=&\int_{\mathcal M} \sqrt{-\gamma}d^{3}x\, \gamma^{ij}(2\phi_{,i}\phi_{,j}+\fr{e^{-4\phi}}{2}\omega_{,i}\omega_{,j}) \nonumber \\
&=&\int_{\mathcal M} \sqrt{-\gamma}d^{3}x\, h_{AB}\Phi^{A}_{,i}\Phi^{B}_{,j}\gamma^{ij}, \label{hfaction}
\eqa
where the fields $\Phi^{A}=\phi\delta^{A}_{1}+\omega\delta^{A}_{2}$ define a mapping between $({\mathcal M},\mathbf{\gamma})$ and a hyperbolic target two-space $V\approx{\mathbb R}^{2}$, with Riemannian metric $h_{AB}=\mbox{diag}(2,e^{-4\phi}/2)$. From Hamilton's equations, it follows that the equations of motion for $\Phi^{A}$ are just the critical points of the functional (\ref{hfaction}), which are given by
\beq
\gamma^{ij}\nabla_{i}\Phi^{A}_{,j}=\gamma^{ij}(\Phi^{A}_{,ij}-\Gamma^{k}_{ij}\Phi_{,k}^{A}+\Gamma^{A}_{BC}\Phi^{B}_{,i}\Phi^{C}_{,j})=0, \label{hmp}
\eeq
which is known as the {\em harmonic map equation}, whose solutions are thus called {\em harmonic maps}~\cite{hmaps}, and play an important role in several different areas of physics~\cite{nuktu74-misner78}.

The stress-energy tensor associated with the harmonic map fields $\Phi^{A}$ is
\beq
T_{ij}[\Phi]:=-\fr{1}{\sqrt{|\gamma|}}\fr{\delta S_{\rm M}}{\delta \gamma^{ij}}=\Phi_{,i}\cdot\Phi_{,j}-\fr{1}{2}\gamma_{ij}\Phi_{,k}\cdot\Phi^{,k}, \label{set}
\eeq
where the dot denotes scalar product in the metric $h_{AB}$ of the target space $V$. One can readily check that $T_{ij}$ is divergence-free:
\beq
\nabla_{i}T^{i}_{j}=\Phi_{,j}\cdot\gamma^{ik}\nabla_{i}\Phi_{,k}=0,
\eeq
where the last equality follows directly from the harmonic map equation. Summarizing, the action for four-dimensional vacuum gravity with one global spacelike KVF of translational type reduces to Eq. (\ref{eha2}), which is the action for $(2+1)$ gravity coupled to harmonic map fields. Since the quotient manifold ${\mathcal M}$ is three-dimensional, the associated spacetime cannot contain any gravitational-wave degrees of freedom (we note, however, that on higher-genus surfaces there are a finite number of Teichm\"{u}ller degrees of freedom, which are absent on ${\mathbb R}^{2}$), and the dynamics must therefore be encoded in the harmonic map fields $\Phi^{A}$ through the $\beta^{i}$ dependence [cf. Eq. (\ref{msplit})], which act as a divergence-free source for the reduced three-dimensional Einstein equations.  $\Box$

Proposition 2: {\em The stress-energy tensor associated with harmonic map fields obeys the dominant energy condition.}

Proof.  The dominant energy condition states that~\cite{wald84}, for all future-oriented timelike vector fields $V^{i}$, the flux vector field $J^{i}\equiv-V^{j}T_{j}^{i}$ is future-oriented and non-spacelike. For harmonic maps $\Phi^{A}$ with stress-energy tensor given by Eq. (\ref{set}), we have
\beq
J^{i}=-\Phi^{,i}\cdot V^{j}\Phi_{,j}+\fr{1}{2}V^{i}\Phi_{,k}\cdot\Phi^{,k}, \label{jay}
\eeq
and thus
\beq
J^{i}J_{i}=\fr{1}{4}(V^{i}V_{i})(\Phi_{,j}\cdot\Phi^{,j})^{2},
\eeq
which is nonpositive for $V^{i}V_{i}<0$, i.e., $J^{i}$ is non-spacelike, for any timelike $V^{i}$. Now, since $V^{i}$ is future-oriented by assumption, $J^{i}$ will be too provided $V^{i}J_{i}\leq0$. From Eq. (\ref{jay}) this condition reads
\beq
-(V^{i}\Phi_{,i})\cdot(V^{j}\Phi_{,j})+\fr{1}{2}(V^{i}V_{i})(\Phi_{,k}\cdot\Phi^{,k})\leq0.
\eeq
An obvious sufficient condition for the inequality to hold is 
\beq
(V^{i}\Phi_{,i})\cdot(V^{j}\Phi_{,j})-(V^{i}V_{i})(\Phi_{,k}\cdot\Phi^{,k})\geq0.
\eeq
On ${\mathcal M}$ introduce locally Gaussian normal coordinates $\{\xi^{i}\}$, wherein $\gamma^{\rm GNC}_{ij}d\xi^{i}d\xi^{j}=-d\tau^{2}+\tilde{\Omega}_{ab}d\xi^{a}d\xi^{b}$, and then rotate the basis vectors such that $V^{i}=\delta^{i}_{\tau}$. The inequality above reads then
\beq
\tilde{\Omega}^{ab}\Phi_{,a}\cdot\Phi_{,b}\geq0,
\eeq
which is evidently satisfied, since $\tilde{\Omega}_{ab}$ and the target-space metric $h_{AB}$ are Riemannian metrics. This completes the proof. $\Box$

Proposition 3 (Ida's Theorem): {\em Let $({\mathcal M},\mathbf{\gamma})$ be a $(2+1)$-dimensional Lorentzian spacetime satisfying the Einstein equations ${\mathbf G}( \gamma)={\mathbf T}$. If ${\mathbf T}$ obeys the dominant energy condition, then there are no apparent horizons in $({\mathcal M},\mathbf{\gamma})$.}

Proof. The idea of the proof---originally due to Hawking for the proof of $S^{2}$ topology of event~\cite{hawking72} and apparent horizons~\cite{hawking73} in asymptotically flat stationary spacetimes, and recently used by Ida~\cite{ida00} in the context of the BTZ black hole---consists in showing that, if an apparent horizon, ${\mathcal A}$, exists {\em and} the dominant energy condition is satisfied, then one could deform ${\mathcal A}$ outward, so as to produce a new closed surface $\hat{\mathcal A}$ just outside ${\mathcal A}$, which is contained in a trapped region, thereby contradicting the ansatz that the former is the outer boundary of a compact trapped region. Ida considers $(2+1)$ gravity with a {\em positive} cosmological constant, $\Lambda>0$; here, we give an outline of the proof for $(2+1)$ gravity coupled to a generic stress-energy tensor with $\Lambda=0$ (albeit subtle, the distinction merits a discussion).

In what follows, we use a three-dimensional analogue of the Newman-Penrose (NP) tetrad formalism~\cite{newman&penrose62}, and have thus changed the signature of the Lorentzian metric $\gamma_{ij}$ to $-1$, to conform with the standard NP construction. We begin by considering a $(2+1)$ foliation ${\mathcal M}\approx {\mathbb R}\times\Sigma$, where $\Sigma$ is a spacelike two-surface. Assume that $\Sigma$ contains a trapped region ${\mathcal T}$---not necessarily simply connected nor compact---and let the outer boundary of ${\mathcal T}$ be a closed curve ${\mathcal C}=\partial{\mathcal T}$, such that ${\mathcal T}$ has the structure of a manifold with a boundary (which is therefore orientable). By definition, ${\mathcal C}$ is an apparent horizon~\cite{wald84b}, characterized by the vanishing of the expansion of future-oriented, outward-pointing null geodesics orthogonal to it. Now let $l^{i}_{\pm}$ be future-oriented null (outgoing/ingoing) vector fields orthogonal to $\Sigma$ at ${\mathcal C}$, and $m^{i}$ be a unit spacelike tangent vector field to ${\mathcal C}$. These vectors form a triad $\{l^{i}_{\pm}, m^{i}\}$, normalized such that
\bqa
l^{\pm}_{i}l^{i}_{\pm}&=&l_{i}^{\pm}m^{i}=0, \;\; l_{i}^{\pm}l^{i}_{\mp}=-m^{i}m_{i}=1, \\
\gamma_{ij}&=&2l^{+}_{(i}l^{-}_{j)}-m_{i}m_{j}.
\eqa
The unit spacelike outward vector orthogonal to ${\mathcal C}$ is $u^{i}=\fr{1}{\sqrt{2}}(l^{i}_{+}-l^{i}_{-})$, which obeys $u^{i}m_{i}=0$, and can always be made to lie on $\Sigma$ by null boosts $l^{i}_{\pm}\rightarrow e^{\pm h}l^{i}_{\pm}$, for some real-valued function $h$. We now deform the curve ${\mathcal C}$ pointwise outwards along the vector field $\xi^{i}=e^{f}u^{i}$ (where $f$ is a real-valued function on $\Sigma$), so as to produce a new closed curve $\hat{\mathcal C}$, satisfying
\beq
\oint_{\hat{\mathcal C}} ds>\oint_{\mathcal C} ds,
\eeq
where $s$ is proper length on $\Sigma$. To keep the deformation confined to $\Sigma$ without destroying the tangency of $m^{i}$ and the orthogonality of $l^{i}_{\pm}$ to $\hat{\mathcal C}$, we require ${\mathcal L}_{\xi} m^{i}={\mathcal L}_{\xi}(l^{i}_{\pm}m_{i})=0$, which leads to
\bqa
\kappa-\tau+\beta-\delta f&=&0, \label{hso1} \\
\nu-\pi-\beta-\delta f&=&0, \label{hso2}
\eqa
where the usual NP notation for the spin coefficients and directional derivatives was adopted:
\bqa
\kappa&=& m_{i}Dl^{i}_{+}, \;\;\;\; \tau= m_{i}\Delta l^{i}_{+}, \;\;\;\; \beta= l^{i}_{-}\delta l_{i}^{+}, \nonumber \\
\nu&=& m_{i}\Delta l^{i}_{-}, \;\;\;\; \pi= m_{i}D l^{i}_{-},  \nonumber \\
D&\equiv& l^{i}_{+}\,^{(3)}\nabla_{i}, \;\;\; \Delta\equiv l^{i}_{-}\,^{(3)}\nabla_{i}, \;\;\; \delta\equiv m^{i}\,^{(3)}\nabla_{i}.
\eqa
The outgoing geodetic expansion associated with $l^{i}_{+}$ is
\beq
\rho:=\,^{(3)}\nabla_{i}l_{+}^{i}=(l^{+}_{i}\Delta+l_{i}^{-}D-m_{i}\delta)l^{i}_{+}=-m_{i}\delta l_{+}^{i},
\eeq
and its change along $\xi^{i}$ is
\beq
{\mathcal L}_{\xi} \rho=\fr{e^{f}}{\sqrt{2}}(D\rho-\Delta\rho), \label{liechange}
\eeq
where the directional derivatives of $\rho$ along $l^{i}_{\pm}$ are given by the NP-like equations
\bqa
D\rho&=&\rho(\epsilon-\rho)-\delta\kappa+(2\beta+\tau+\pi)\kappa+\varphi_{++}, \label{npe1} \\
\Delta\rho&=&\rho(\mu-\gamma)-\delta\tau+\tau^{2}+\kappa\nu+\varphi_{+-}, \label{npe2}
\eqa
where
\bqa
\epsilon&=&l_{i}^{-}Dl^{i}_{+}, \;\;\;
\gamma=l^{+}_{i}\Delta l^{i}_{-}, \;\;\;
\mu=m_{i}\delta l^{i}_{-}, \nonumber \\
\varphi_{++}&=&R_{ijkl}l^{i}_{+}m^{j}l^{k}_{+}m^{l}=-R_{ik}l^{i}_{+}l^{k}_{+}, \nonumber \\
\varphi_{+-}&=&R_{ijkl}l^{i}_{+}m^{j}l^{k}_{-}m^{l}=R_{ik}l^{i}_{+}l^{k}_{-}-\fr{R}{2}. \label{last}
\eqa
Since $\rho=0$ on $\hat{\mathcal C}$ by construction, Eqs. (\ref{hso1}), (\ref{hso2}), and (\ref{liechange})-(\ref{last}) yield 
\beq
\sqrt{2}e^{-f}{\mathcal L}_{\xi}\rho=\delta(\beta-\delta f)-(\tau-\kappa)^{2}-R_{ij}(l^{i}_{+}l^{j}_{+}+l^{i}_{+}l^{i}_{-})+\fr{R}{2}. \label{liefinal}
\eeq
The second term is manifestly negative~\footnote{This term only vanishes if $\tau=\kappa$. In our construction $l^{i}_{+}$ is a null geodesic vector field, whence $\kappa\equiv0$, which forces $\tau$ to vanish. The normalization for the triad $\{l^{i}_{\pm},m^{i}\}$, together with the fact that $l^{i}_{+}$ is assumed to be geodesic, imposes eight conditions (seven algebraic, and one differential equation) on the nine functional degrees of freedom of the triad vector fields, thereby reducing the system to {\em only one} functional degree of freedom for the $m^{i}$ and $l^{i}_{-}$ vector fields. The condition $\tau=0$ then fully determines the triad, and hence the curve ${\mathcal C}$. Clearly, any arbitrarily small perturbation of such curve will yield another curve which does {\em not} satisfy $\tau=0$. Accordingly, for a generic closed curve ${\mathcal C}$ we have $\tau\neq0$.}, and the first term can always be made to vanish on $\hat{\mathcal C}$ by appropriate choice of $f$, e.g., by imposing $\beta=\delta f$, which leads to a first-order linear PDE for $f$ on $\Sigma$, or by parametrizing $\hat{{\mathcal C}}$ by proper length $s$ and defining $f:=\int_{s} \beta ds-c_{0}s, \; c_{0}\in{\mathbb R}$ [which identically satisfies $\delta(\beta-\delta f)=0$] . Thus, a sufficient condition for the right-hand-side of Eq. (\ref{liefinal}) to be strictly negative is that the last two terms are nonpositive. Einstein's equations, ${\mathbf G}={\mathbf T}$, together with the assumption that ${\mathbf T}$ obeys the dominant energy condition, imply (by continuity for null vectors)
\beq
{\mathbf G}(l_{+},l_{+})=R_{ij}l^{i}_{+}l^{j}_{+}\geq0, \;\;
{\mathbf G}(l_{+},l_{-})=R_{ij}l^{i}_{+}l^{j}_{-}-\fr{R}{2}\geq0.
\eeq
Therefore, the right-hand-side of Eq. (\ref{liefinal}) is strictly negative~\footnote{This equation can only vanish identically if, in addition to taking a highly non-generic curve ${\mathcal C}$ which satisfies $\kappa=\tau=0$, we also have $R_{ij}l^{i}_{+}l^{j}_{+}=-R_{ij}l^{i}_{+}l^{j}_{-}+R/2=0$ everywhere along the curve. However, $l^{i}_{\pm}$ are fully determined at this point, and thus these two conditions will not in general (if at all) be satisfied. The right-hand-side of Eq. (\ref{liefinal}) is therefore strictly negative for a generic curve ${\mathcal C}$.}, i.e., future-oriented outgoing null geodesics orthogonal to $\hat{\mathcal C}$ are {\em converging} (negative expansion). But this means that $\hat{\mathcal C}$ is contained in a trapped region, which contradicts the assumption that the former is an outer marginally trapped surface. Hence, $({\mathcal M},\mathbf{\gamma})$ cannot contain apparent horizons. $\Box$

Theorem: {\em Let $(M,\mathbf{g})$ be a four-dimensional Lorentzian spacetime obeying the vacuum Einstein equations ${\mathbf G}(\mathbf{g})=0$. If $(M,\mathbf{g})$ admits a global spacelike $G_{1}$ group of translational isometries, then it cannot contain $G_{1}$-invariant apparent horizons.}

We first note that, the translational symmetry precludes apparent horizons from being homeomorphic to $S^{2}$---the standard topology in asymptotically flat spacetimes---since one can always continuously deform any such surface along the symmetry direction, whereby the property of outer boundary of a compact region is lost: Take a spacelike hypersurface $^{(3)}\Sigma\supset{\mathcal R}$, where ${\mathcal R}\approx S^{2}$ is assumed to be an apparent horizon. Let $^{(3)}\Sigma={\mathbb R}\times\,^{(2)}\Sigma$ be foliated by a family of two-surfaces $^{(2)}\Sigma_{z}\approx{\mathbb R}^{2}$, where each $z\in{\mathbb R}$ denotes a given slice.Translational invariance means that any two such slices are isometric. Take then a slice $^{(2)}\Sigma_{z_{1}}$ such that $^{(2)}\Sigma_{z_{1}}\cap{\mathcal R}={\mathcal C}\approx S^{1}$, and consider the spacelike hypersurface $^{(3)}\Xi=[0,a]\times\,^{(2)}\Sigma_{z_{1}}$, where $a\in{\mathbb R}^{+}\backslash\{0\}$. The manifold
$$^{(3)}\tilde{\Sigma}=\{(-\infty,z_{1}]\times\,^{(2)}\Sigma_{z_{1}}\}\cup\,^{(3)}\Xi\cup\{[z_{1},+\infty)\times\,^{(2)}\Sigma_{z_{1}}\}$$
is geometrically identical to $^{(3)}\Sigma$ and has an ``extended'' apparent horizon which contains ${\mathcal R}$ by construction---this contradicts the assumption that ${\mathcal R}$ is the {\em outer} boundary of a compact trapped region. Similar arguments can also be given to rule out topological $S^{1}\times{\mathbb R}$ surfaces which are not $G_{1}$-invariant~\footnote{Take a given spacelike hypersurface $^{(3)}\Sigma$ and consider the foliation $^{(2)}\Sigma_{z}: \{x^{3}=z=\mbox{const.}\}$, where all $^{(2)}\Sigma_{z}$ slices are isometric by $G_{1}$ invariance. Suppose there is a {\em non-}$G_{1}$-invariant topological $S^{1}\times{\mathbb R}$ trapped surface (not necessarily an apparent horizon) in $^{(3)}\Sigma$, and take two different slices $^{(2)}\Sigma_{z_{1}}$, $^{(2)}\Sigma_{z_{2}}$; on each two-slice one would obtain {\em different} topological circles (geometric, for the case of surfaces of revolution around the $x^{3}$-axis). However, such circles are entirely determined by the metric---via the geodesic equation and the requirement that the expansion of outgoing null geodesic congruences orthogonal to the surface vanishes thereon---which, by $G_{1}$-invariance, must be the {\em same} on both slices. This is a contradiction, and thus such $x^{3}$-dependent surfaces cannot exist.}.

We shall therefore take apparent horizons in ${\mathbb R}$-symmetric spacetimes to be  {\em $G_{1}$-invariant topological $S^{1}\times{\mathbb R}$ (where the ${\mathbb R}$ factor corresponds to the orbits of the KVF) spacelike two-surfaces, which are outer marginally trapped, and are the outer boundary of a (non-compact) trapped region along the spacelike two-sector of the quotient Lorentzian spacetime induced by the orbits of the KVF.}

Proof: Consecutive application of Propositions 1 thru 3 implies that there are no apparent horizons in the quotient three-dimensional spacetime $({\mathcal M}\approx M/{\mathbb R},\gamma)$, which arises from dimensional reduction of vacuum $(3+1)$ gravity under $G_{1}$ actions. We must now relate the absence of apparent horizons in the $(2+1)$ spacetime to that in the original $(3+1)$ picture. The proof proceeds by {\em reductio ad absurdum}: assume that an apparent horizon exists in $(M,\,^{(4)}g_{\mu\nu})$, and then show that this implies that the reduced $(2+1)$ spacetime also contains an apparent horizon, which contradicts Proposition 3. Let us then assume that there is a spacelike hypersurface $^{(3)}\Sigma$ which contains an apparent horizon, i.e., a two-surface $^{(2)}{\mathcal A}\approx S^{1}\times{\mathbb R}$, which is the outer boundary of a trapped region, and satisfies 
\beq
\left[^{(4)}\nabla_{\mu}\,^{(4)}l^{\mu}\right]_{^{(2)}\!{\mathcal A}}=0, \label{omt}
\eeq 
where $^{(4)}l^{\mu}$ is the vector field tangent to future-oriented outgoing null geodesics orthogonal to $^{(2)}\!{\mathcal A}$. Now take a spacelike two-surface $^{(2)}\Sigma \subset\,^{(3)}\Sigma$ transverse to the orbits of the KVF, so that $$^{(2)}\Sigma\,\cap\,^{(2)}{\mathcal A}={\mathcal C}\approx S^{1},$$ where ${\mathcal C}$ has tangent vector field $^{(3)}t^{i}$ and future-oriented outgoing null normal $^{(3)}l^{i}$, with respect to the Lorentzian three-metric $^{(3)}g_{ij}$ on the manifold given by the Cartesian product ${\mathbb R}\times\,^{(2)}\Sigma$. By construction, ${\mathcal C}$ is the outer boundary of a trapped region in the reduced spacetime~\footnote{If ${\mathcal C}$ was interior to the actual trapped region, simply by taking the product ${\mathbb R}\times\,^{(2)}\Sigma$ one would obtain a $(2+1)$ spacetime with a {\em trapped} topological cylinder which coincides with the original apparent horizon one started with, which is a contradiction. If ${\mathcal C}$ were outside the true outer boundary, an analogous procedure would lead to the contradiction that the apparent horizon one started with was {\em untrapped}; specifically, there would be a region with topology $[0,a]\times S^{1}\times{\mathbb R}$, whose outer boundary coincides with the original ``apparent horizon'', which is untrapped.}, but one still needs to require that the projection of $^{(4)}l^{\mu}$ onto the quotient manifold,
\beq
l^{i}_{\perp}\equiv\perp^{i}_{\mu}\,^{(4)}l^{\mu}=(\delta_{\mu}^{i}-\,^{(4)}\xi_{\mu}\,^{(4)}\xi^{i})\,^{(4)}l^{\mu},
\eeq
be collinear with $^{(3)}l^{i}$ {\em and} has vanishing divergence. This leads to the conditions
\bqa
l^{i}_{\perp}l^{\perp}_{i}&=&0, \label{co1} \\
l^{i}_{\perp}\,^{(3)}t_{i}&=&0, \label{co2} \\
^{(3)}\nabla_{i}l^{i}_{\perp}&=&0. \label{co3}
\eqa
What is fixed, and what are the variables in the equations above? Equation (\ref{co1}) reads
\bqa
&&[e^{-2\phi}(-\tilde{N}^{2}+\tilde{\sigma}_{ab}\tilde{N}^{a}\tilde{N}^{b})+e^{2\phi}\beta^{2}_{0}]\left(l^{t}_{\perp}\right)^{2} \nonumber \\
&&+2(e^{-2\phi}\tilde{\sigma}_{ab}\tilde{N}^{b}+e^{2\phi}\beta_{0}\beta_{a})l^{a}_{\perp}l^{t}_{\perp} \nonumber \\
&&+ (e^{-2\phi}\tilde{\sigma}_{ab}+e^{2\phi}\beta_{a}\beta_{b})l^{a}_{\perp}l^{b}_{\perp}=0.
\eqa
Equation (\ref{co2}) has a similar form. The components $l^{i}_{\perp}$ and $^{(3)}t^{i}$ are automatically given by the choice of slicing surface $^{(2)}\Sigma$, and the objects $\{\phi,\beta_{0},\beta_{a},\tilde{\sigma}_{ab}\}$ are determined by the field equations (either in full four-dimensional form, or in the equivalent reduced formulation). Equations (\ref{co1})--(\ref{co2}) form thus a coupled system of second-degree polynomial equations for the three variables $\{\tilde{N},\tilde{N}^{a}\}$, which one may solve for $\tilde{N}^{a}$ for a given $\tilde{N}$. Condition (\ref{co3}) may be written as
\bqa
^{(3)}\nabla_{i}l^{i}_{\perp}&=&\fr{1}{\sqrt{|^{(3)}g|}}\left(\sqrt{|^{(3)}g|}l^{i}_{\perp}\right)_{,i} \nonumber \\
&=&\left(\ln\sqrt{|^{(3)}g|}\right)_{,i}l^{i}_{\perp}+l^{i}_{\perp,i}=0.
\eqa
The second term is known, and the first term contains only first derivatives of $\tilde{N}$, i.e., we have a linear first-order PDE of gradient type for $\tilde{N}$, wherein existence and uniqueness follow from standard linear PDE theory~\cite{taylor96}. Geometrically, the lapse $\tilde{N}$ and shift $\tilde{N}^{a}$ control the embedding of the spacelike two-surface $^{(2)}\Sigma$ in the $(2+1)$ spacetime (the intrinsic geometry of $^{(2)}\Sigma$ is given by $\tilde{\sigma}_{ab}$, and is of course independent of such embedding), whereby the full three-metric is determined. It is precisely this gauge freedom which allows us to construct a $(2+1)$ spacetime whose metric is such that Eqs. (\ref{co1})--(\ref{co3}) are satisfied, whence ${\mathcal C}$ is an apparent horizon in the reduced spacetime. But this contradicts Proposition 3, and thus apparent horizons cannot exist in $(M,\,^{(4)}g_{\mu\nu})$. $\Box$

By means of example, we work out explictly the case where the KVF is hypersurface-orthogonal ($\beta_{a}=\beta_{0}=0$) and has constant norm ($\phi=\phi_{\rm c}=\mbox{const.}$):
\beq
^{(4)}ds^{2}=g_{\mu\nu}dx^{\mu}dx^{\nu}=e^{-2\phi_{\rm c}}\gamma_{ij}dx^{i}dx^{j}+e^{2\phi_{\rm c}}(dx^{3})^{2}.
\eeq
Note that $|g|\equiv\mbox{det}(g_{\mu\nu})=e^{-4\phi}\mbox{det}(\gamma_{ij})\equiv e^{-4\phi}|\gamma|$. Let $^{(3)}l^{i}$ be the vector field tangent to future-oriented outgoing null geodesics orthogonal to a given closed spacelike one-curve in ${\mathcal M}$. In terms of the four-dimensional coordinates we have 
\beq
^{(4)}l^{\mu}=\,^{(3)}l^{i}\delta_{i}^{\mu}\equiv l^{i}_{\perp},
\eeq
which automatically satisfies Eqs. (\ref{co1})--(\ref{co2}). The four-divergence of $^{(4)}l^{\mu}$ is then
\bqa
^{(4)}\nabla_{\mu}\,^{(4)}l^{\mu}&=&\fr{1}{\sqrt{|g|}}(\sqrt{|g|}\,^{(4)}l^{\mu})_{,\mu} \nonumber \\
&=&\fr{1}{\sqrt{|\gamma|}}(\sqrt{|\gamma|}l^{i}_{\perp})_{,i}\equiv\, ^{(3)}\nabla_{i}l^{i}_{\perp}.
\eqa
Since $^{(4)}\!\nabla_{\mu}\,^{(4)}l^{\mu}=0$ by assumption, Eq. (\ref{co3}) is also automatically satisfied. 

This theorem constitutes thus compelling evidence towards the validity of the hoop conjecture, which puts forward a necessary and sufficient condition for horizon formation in general relativity: {\em horizons form if and only if mass is sufficiently compacted in all three spatial directions}. Our result explains---in the context of vacuum gravity---the ``only if'' part of the conjecture, by showing that if mass cannot be compacted along one spacelike direction, then the spacetime cannot contain apparent horizons. This no-horizon property of vacuum gravity with a translational spacelike KVF is not a mere geometrical artifact; rather, it is a genuine feature of the theory, enforced by the field equations. It should be clear that the inclusion of a {\em positive} cosmological constant leaves our conclusions unchanged. The result holds for {\em any} vacuum metric with translational symmetry, wherein all known axial and cylindrical solutions---for which the absence of apparent horizons has been explicitly demonstrated---are included. We further remark that, unlike several existing trapped-surface results, which resort to larger isometry groups [e.g., $SO(3)$], special choice of spacelike slicing (e.g., maximal), and restrictions on the initial data (e.g., time-symmetric), this result is free from all such limitations, and relies solely on local properties of null geodesic congruences, which are insensitive to the details of the metric and initial data, but {\em depend crucially on the dimensionality of the spacetime}: in three spacetime dimensions, terms that would otherwise (in four dimensions) be present in the NP-like equations (\ref{npe1})-(\ref{npe2}), are now absent, which is precisely what allows for the difference of these two equations to be strictly negative, whereby Proposition 3 is proved. 

Finally, we point out that this result also has potential implications for cosmic censorship for vacuum gravity with the chosen KVF. To the author's best knowledge, there are no large data global existence results for the vacuum Einstein equations with one spacelike KVF~\cite{bcm97}, which can thus---at least {\em a priori}---develop nonspacelike singularities. Any such singularity would be at least locally naked, since the spacetime is free of apparent horizons, thus in clear violation of the strong cosmic censorship conjecture. The implications for weak cosmic censorship are less clear, since the former requires an event horizon, the definition of which for non-asymptotically-flat spacetimes is lacking. Although it seems likely that event horizons (defined in a suitable sense) cannot exist either---the no-horizon proof in this paper is {\em independent} of a particular spacelike foliation choice, and it is difficult to imagine a spacetime that is free of closed trapped regions for {\em every} possible slicing and yet contains an event horizon---because of ambiguities with the definition of the former we shall not pursue this discussion here.

\begin{acknowledgments}
I am indebted to Vince Moncrief for many valuable conversations and comments, and for his reading of an early version of the manuscript. I also thank Daisuke Ida for private communications. This work was supported in part by FCT Grant SFRH-BPD-5615-2001 and by NSF Grant PHY-0098084.
\end{acknowledgments}

\end{document}